\newcommand*{\ATLASLATEXPATH}{}
\documentclass[UKenglish,texlive=2011,txfonts,cernpreprint, texmf]{\ATLASLATEXPATH atlasdoc}

\usepackage{\ATLASLATEXPATH atlaspackage}
\usepackage{overpic}
\usepackage{\ATLASLATEXPATH atlasbiblatex}
\usepackage{units}
\usepackage{\ATLASLATEXPATH atlascontribute}

\usepackage{\ATLASLATEXPATH atlasphysics}

\def\emu{\mbox{$e^{\pm}\mu^{\mp}$}}
\def\coscos{\mbox{$\cos\theta_1\cdot\cos\theta_2$}}

\AtlasTitle{Measurement of the correlations between the polar angles of leptons from top quark decays in the helicity basis at $\sqrt{s}=7$~TeV using the ATLAS detector}

\author{The ATLAS Collaboration}

\AtlasRefCode{TOPQ-2013-10}

\PreprintIdNumber{CERN-PH-EP-2015-241}

\AtlasJournal{Phys.\ Rev.\ D. \-RC }

\AtlasAbstract{
A measurement of the correlations between the polar angles of leptons from the decay of pair-produced
$t$ and $\bar{t}$ quarks in the helicity basis is reported, using
proton-proton collision data collected by the ATLAS detector at the LHC\@.
The dataset corresponds to an integrated luminosity of 4.6~fb$^{-1}$ at
a center-of-mass energy of $\sqrt{s}=7$~\TeV\ collected during 2011.
Candidate events are selected in the dilepton topology with large missing transverse momentum and at least two jets.
The angles $\theta_1$ and $\theta_2$ between the charged leptons and the direction of motion of the parent quarks
in the \ttbar\ rest frame are sensitive to the spin information,
and the distribution of $\coscos$ is sensitive to the spin correlation between the $t$ and $\bar{t}$ quarks.
The distribution is unfolded to parton level and compared to the
next-to-leading order prediction.
A good agreement is observed.
}

\AtlasCoverSupportingNote{ATL-COM-PHYS-2012-1065}{https://cds.cern.ch/record/1462276}
\AtlasCoverCommentsDeadline{10 September 2015}

\AtlasCoverAnalysisTeam{Liming Chen, Simon Head, James Howarth, Kevin Kroeninger, Boris Lemmer, Bo Li,Tom McLaughlan, Emmanuel Monnier, Tom Neep, Mark Owen, Yvonne Peters, Arnulf Quadt, Christian Schwanenberger, Elizaveta Shabalina, Peter Watkins, John Wilson, Chengguang Zhu}

\AtlasCoverEdBoardMember{Antonio Onofre (chair), Ian Brock, Giuseppe Salamanna, Markus Wobisch}

\AtlasCoverEgroupEditors{atlas-topq-2013-10-editors@cern.ch}

\AtlasCoverEgroupEdBoard{atlas-topq-2013-10-editorial-board@cern.ch}

\hypersetup{pdftitle={ATLAS draft},pdfauthor={The ATLAS Collaboration}}

\begin{document}

\maketitle

\section{Introduction}
\label{sec:intro}

After the discovery of the top quark in 1995 at the Tevatron proton--antiproton
collider~\cite{bib:topFoundD0, bib:topFoundCDF}, a new era of top quark precision
measurements began in 2010 with the start of the Large Hadron Collider (LHC) at CERN
at $\sqrt{s}=7~\TeV$\@.
Due to its short lifetime the top quark decays before hadronization.
This implies that top quarks can be studied as bare quarks
and the spin information of the top quark can be deduced from the angular distributions of its decay products.
At the LHC, $\ttbar$ production is dominated by gluon fusion with a smaller contribution from $q\bar{q}$ annihilation.
However, many scenarios of physics beyond the Standard Model (SM) predict different spin correlations.
For example, the measured spin correlation may differ from the SM
if $\ttbar$ production from $q\bar{q}$ annihilation was enhanced by the top quark coupling
to Higgs or extra gauge bosons~\cite{Arai:2008qa,bib:hep-ph-9911288,PhysRevD.58.114031,epjc-66-261},
or if the top quark decayed into a scalar charged Higgs boson and a $b$-quark ($t \rightarrow H^+ b$)~\cite{Abazov:2009ae}.

Both the CDF and D\O\ collaborations have performed measurements of the spin correlation~\cite{Aaltonen:2010nz,Abazov:2011ka,Abazov:2011qu, Abazov:2011gi,PhysRevD.87.011103}
at the Tevatron where $\ttbar$ production via $q\bar{q}$ annihilation dominates.
In addition, it has been measured
at the LHC by both the ATLAS~\cite{TOPQ-2011-11,TOPQ-2013-01,TOPQ-2014-07}
and CMS experiments~\cite{CMS-TOP-13-003}.
The different production mechanisms and center-of-mass energies make the measurements
of the spin correlation at the two colliders complementary~\cite{Bernreuther:2004jv}.
The results obtained from these analyses are all consistent with the SM prediction.

In this paper the decay $\ttbar \rightarrow W^+ W^- b \bar{b} \rightarrow \ell^+ \nu \ell^- \bar{\nu} b \bar{b}$ is
used to measure the following differential distribution, 
which is related to the spin correlation of the $\ttbar$ system~\cite{Bernreuther:2010ny}

\begin{equation}
\frac{1}{N} \frac{ d^2 N}{d\cos \theta_1 d\cos \theta_2} =
\frac{1}{4} \left( 1 + B_1 \cos \theta_1 + B_2 \cos \theta_2 - C_{\textrm{helicity}}  \coscos \right),
\label{eq:spinAngDisTTbar}
\end{equation}

\noindent
where $\theta_1 \left( \theta_2 \right)$ is the angle between the momentum direction of the
charged lepton from the $t \left( \bar{t} \right)$
decay in the $t \left( \bar{t} \right)$ rest frame and the $t \left( \bar{t} \right)$ momentum direction in the
$\ttbar$ center-of-mass frame.
This is commonly referred to as the helicity basis.
The helicity basis is not the only possibility and other bases are discussed in Ref.~\cite{Bernreuther:2004jv}.
The top quark polarization parameters $B_1$ and $B_2$
are two orders of magnitude smaller than $C_{\textrm{helicity}}$ at next-to-leading-order (NLO)~\cite{Bernreuther:2013aga}.
In Ref.~\cite{TOPQ-2012-22} the polarization is measured to be
$-0.035 \pm 0.040$ for the CP-conserving scenario, compatible
with the measurement from CMS~\cite{CMS-TOP-13-003} and the SM expectation~\cite{PhysRevD.84.074034}.
Consequently they are set to zero in this study.
This analysis uses the measured distribution of $\coscos$: 
it can be shown~\cite{Bernreuther:2010ny} that the mean of the distribution is proportional to the coefficient $C_{\textrm{helicity}}$ parameterizing the strength of the spin correlation.

Candidate events are selected with two isolated charged leptons
and at least two jets
in the final state, including a requirement to enhance the selection of jets originating from $b$-quarks.
The $t$ and $\bar{t}$ are
reconstructed
using kinematic information from the event and invariant mass constraints.
The distribution of $\coscos$ at reconstruction level is obtained.
Building upon previous studies the non-$t\bar{t}\quad$  backgrounds are subtracted and the distribution is
unfolded to parton level using an iterative Bayesian technique.
The parton-level distribution can be compared directly with the theoretical prediction 
without the need for templates derived from simulation.

\section{ATLAS detector and data samples}
\label{sec:detector}

This analysis makes use of an integrated luminosity of 4.6~\ifb~\cite{DAPR-2011-01} of proton--proton
collision data at a center-of-mass energy of 7~\TeV, collected by the
ATLAS detector at the LHC during 2011\@.
The ATLAS detector~\cite{PERF-2007-01,PERF-2010-01} covers nearly the
entire solid angle\footnote{ATLAS uses a right-handed coordinate system with its origin
at the nominal interaction point (IP) in the center of the detector and the $z$-axis along the beam pipe. The $x$-axis points from the
IP to the center of the LHC ring, and the $y$-axis points upward. Cylindrical coordinates $(r,\phi)$ are used in the transverse plane,
$\phi$ being the azimuthal angle around the $z$-axis. The pseudorapidity is defined in terms of the polar angle
$\theta$ as $\eta=-\ln\tan(\theta/2)$. 
Angular distance $\Delta R$ is defined as $\sqrt{ (\Delta \phi)^2 + (\Delta \eta)^2}$ where
$\Delta \phi$ and $\Delta \eta$ are the difference of azimuthal angle and pseudorapidity, respectively.} around the collision point. 
It consists of an inner tracking detector (ID) covering $|\eta| < 2.5$, and comprising a silicon
pixel detector, a silicon microstrip detector,
and a transition radiation tracker. The ID is surrounded
by a thin superconducting solenoid providing a
2~T magnetic field, followed by a liquid-argon electromagnetic
sampling calorimeter (LAr) with high granularity. 
An iron/scintillator tile calorimeter provides hadronic energy 
measurements in the central region ($|\eta| < 1.7$).
The endcap and forward regions are instrumented with LAr
calorimeters for electromagnetic (EM) and hadronic energy
measurements up to $|\eta| = 4.9$. The calorimeter system is surrounded
by a muon spectrometer (MS) with high-precision tracking 
chambers covering $|\eta| < 2.7$ and separate trigger chambers covering $|\eta| < 2.4$.
The magnetic field is provided by a barrel and two endcap 
superconducting toroid magnets.
A three-level trigger system is used to select 
events with high-\pt\
leptons for this analysis~\cite{Aad:2012xs}. 
 
Monte Carlo (MC) samples are produced for signal and background estimation.
The SM $\ttbar$ signal events are modeled using the {\sc MC@NLO} v4.01 generator~\cite{Frixione:2003ei}.
Top quarks and the subsequent $W$ bosons are decayed conserving the spin correlation information. 
The decay products are interfaced with {\sc Herwig} v6.520~\cite{xsec_23}, which hadronizes the $b$-quarks and $W$ boson decay products, 
and with {\sc Jimmy}~\cite{jimmy} to simulate multiparton interactions. 
The top quark mass is set to $172.5$~\GeV\ and the 
CT10 parton distribution functions (PDF)~\cite{PhysRevD.82.074024} are used.
The \ttbar\ signal is normalized to \mbox{ $\sigma_{\ttbar} = 177^{+10}_{-11}$~pb},
calculated at next-to-next-to-leading-order (NNLO) in QCD including
resummation of next-to-next-to-leading logarithmic soft gluon terms with
{\sc Top++} v2.0~\cite{Cacciari:2011hy,Baernreuther:2012ws,Czakon:2012zr,Czakon:2012pz,Czakon:2013goa,Czakon:2011xx}.
The single-top-quark background arises from associated $Wt$ production, when both the $W$ boson 
from the top quark and the $W$ boson from the hard interaction decay leptonically.
Events are generated using the {\sc MC@NLO} generator~\cite{Frixione:2005vw} 
and normalized to
\mbox{ $\sigma_{Wt} = 15.7\pm1.2$~pb}
from the approximate NNLO calculation in Ref.~\cite{Kidonakis:2010ux}.

Drell--Yan $Z+$jets events are generated using the {\sc Alpgen} v2.13~\cite{Alpgen} 
generator including leading-order (LO) matrix elements with up
to five additional partons.  
The MLM matching scheme is used to remove overlaps 
between the $n$ and $n+1$ parton samples~\cite{Alpgen}.
The CTEQ6L1 PDF~\cite{CTEQ66,JHEP2002-0207} set is used and the cross-section is normalized
to the NNLO prediction~\cite{Melnikov:2006kv}.  Parton showering and hadronization are modeled by {\sc Herwig}
and the underlying event is simulated by {\sc Jimmy}\@.
The diboson backgrounds ($WW$, $WZ$, $ZZ$) are generated using {\sc Alpgen} 
interfaced to {\sc Herwig}, and
make use of the MRST LO PDF set~\cite{Sherstnev:2007nd}.
They are all normalized to the theoretical predictions at NLO~\cite{Campbell:1999ah}.

All MC samples use a {\sc GEANT4} based simulation~\cite{xsec_13,SOFT-2010-01} to
model the ATLAS detector and the same reconstruction as used in data.
During the 2011 data-taking period the average number of simultaneous $pp$
interactions per beam crossing (pileup) at the beginning of a fill of the LHC increased from 6 to 17.
For each MC process, pileup is overlaid using simulated minimum-bias 
events from the {\sc Pythia} generator. The number of additional $pp$ interactions is 
reweighted to the number of interactions observed in data.
Additional small corrections are made to the simulation to ensure that it describes
the data well in terms of efficiencies and momentum or energy scales for the various objects used.

While all other backgrounds are based on MC simulation, 
the background arising from misidentified and nonprompt leptons (referred to as ``fake leptons'' in the figures and tables)
is determined using a data-driven technique known as the matrix method
~\cite{ATLAS-CONF-2014-058}.

\section{Event selection and reconstruction}
\label{sec:SelReco}

\subsection{Event selection}

Candidate events are selected in the dilepton topology,
referred to as the $\ee$, $\mumu$, and $\emu$ channels, according to the flavors of the two leptons.
The full object and event selection are the same as described in Ref.~\cite{TOPQ-2013-01},
with the additional requirement that at least one $b$-jet is identified.
The analysis requires events selected by an inclusive single-lepton($e$ or $\mu$) trigger~\cite{PERF-2011-02}.
The primary vertex with highest $p_{\rm T}^2$ is taken as the primary vertex of the event if it has at least five associated tracks,
with  $\pT > 400$~\MeV\ per track,
consistent with the $x, y$ profile of the beam, and the other vertices are not considered.

Electron candidates are reconstructed using energy deposits in the EM calorimeter
associated with reconstructed tracks in the ID~\cite{PERF-2013-03}\@.
Muon candidate reconstruction makes use of tracking in the MS and ID~\cite{PERF-2011-01}\@.
Both the electron and muon candidates have isolation criteria applied as in Ref.~\cite{TOPQ-2013-01}
and are matched to a triggered object.
Jets are reconstructed with the anti-$k_t$ algorithm~\cite{antikt}
with a radius parameter $R = 0.4$, starting from
energy deposits in clusters of adjacent calorimeter cells.
The missing transverse momentum magnitude \met is
reconstructed from the vector sum of all
calorimeter cell energies associated with topological clusters
with $|\eta| < 4.5$~\cite{Aad:2012re}. Contributions from the calorimeter
energy clusters matched with a reconstructed lepton
or jet are corrected to the corresponding energy scale.
A term accounting for the \pT\ of any selected muon is included in the \met\ calculation.
The following kinematic requirements are made:

\begin{itemize}

\item Electron candidates are required to have $\pt > 25$~\gev\
and $|\eta| < 2.47$, excluding electrons from the transition
region between the barrel and endcap calorimeters defined by $1.37 < |\eta| < 1.52$.
Muon candidates are required to have $\pt > 20$~\gev\ and $|\eta| < 2.5$.

\item Events must have at least two jets with $\pt > 25$~\gev\
and $|\eta| < 2.5$.
Jets associated with large energy deposits from additional $pp$
interactions are suppressed by requiring that the \pt\ sum of the reconstructed
tracks matched to both the jet and the primary vertex is at least 75\%
of the total \pT\ sum of all tracks associated with the jet.
This quantity is referred to as the jet vertex fraction (JVF)~\cite{ATLAS-CONF-2014-018}.
Jets satisfying $\pT > 50~\GeV$ are always accepted and jets having no associated tracks are also accepted.
The jet candidate closest to an accepted electron candidate is removed if it is within $\Delta R < 0.2$.
Finally, electron and muon candidates that lie within a cone of $\Delta R = 0.4$ around an accepted jet are removed.

\item Events must have exactly two oppositely charged
lepton candidates (\ee, \mumu, \emu).

\item At least one of the selected jets must be identified as originating from a $b$-quark ($b$-tagged) using the
multivariate discriminant MV1~\cite{ATLAS-CONF-2012-043}, which
uses impact parameter and secondary vertex information. The chosen MV1 working point corresponds
to an average $b$-tagging efficiency of 70\% for $b$-jets in simulated $t\bar{t}$ events.

\item Events in the \ee\ and \mumu\ channels are required to have
$m_{\ell\ell} > 15$~\gev\ to exclude regions not well described by the MC simulation and to
remove contributions from $\Upsilon$ and
$J/\psi$ production.

\item Events in the \ee\ and \mumu\ channels must satisfy $\met > 60$~\gev\
to suppress the background from $Z/\gamma^*$+jets.
In addition, $m_{\ell\ell}$ must differ by at least 10~\gev\ from
the $Z$ boson mass to further suppress the $Z/\gamma^*$+jets
background.

\item For the \emu\ channel, no \met\ or $m_{\ell\ell}$ cuts are applied.
In this case, the remaining background from
$Z/\gamma^* (\ra \tau\tau)$+jets production is further suppressed by requiring
that the scalar sum of the \pt\ of all selected jets and leptons is greater than 130~\gev.

\end{itemize}

\subsection{Topology reconstruction method}

In order to reconstruct the distribution of \coscos, the $t$ and $\bar{t}$ quarks must
be reconstructed from their decay products.
The momenta of the two neutrinos
from the $W$ boson decays in dilepton final states cannot
be measured but can be inferred from the measured missing transverse
momentum in the event. Since only the sum of the missing transverse momenta
of the two neutrinos is measured as $\met$, the system is underconstrained.
The kinematic information about the $b$-jets,
leptons, and the missing transverse momentum are used in
six independent equations describing the kinematic properties of $t$ and $\bar{t}$ decays.
The equations describing the kinematic constraints are:

\begin{align}
  p_{\nu, x} + p_{\bar{\nu}, x} &= E_{x}^{\text{miss}}, \nonumber \\
  p_{\nu, y} + p_{\bar{\nu}, y} &= E_{y}^{\text{miss}}, \nonumber  \\
  (p_{\ell^{-}} + p_{\bar{\nu}})^{2} &= m_{W^{-}}^{2}, \label{eq:sixFun} \\
  (p_{\ell^{+}} + p_{\nu})^{2} &= m_{W^{+}}^{2},\nonumber  \\
  (p_{W^{-}} + p_{\bbar})^{2} &= m_{\tbar}^{2}, \nonumber  \\
  (p_{W^{+}} + p_{b})^{2} &= m_{t}^{2}, \nonumber
\end{align}

\noindent
where $E_{x}^{\text{miss}}$ and $E_{y}^{\text{miss}}$ represent the missing
momentum along the $x$- and $y$-axes, $p_{\ell^{+}}$
and $p_{\ell^{-}}$ ($p_{b}$ and $p_{\bbar}$) are the
four-momenta of the two charged leptons (two $b$-jets), and
$m_{W}$ and $m_{t}$ are the masses of the $W$ boson and top quark.
The reconstruction algorithm requires the kinematic information for exactly two of the selected jets.
Jets are ranked primarily by whether they are $b$-tagged or not, and then by descending \pT.
The two highest-ranked jets are used in the reconstruction method.

Each selected event has two possible $b$--$\ell$ pairings.
The pairing with the lower invariant mass is first
considered for the $\ttbar$ reconstruction.
If no solution is found, the top quark mass is varied from the nominal value in steps of 1.5~\GeV\ until a solution is found
or the limits of 157.5~\GeV\ and 187.5~\GeV\ are reached.
If it is still not possible to solve Eq.(~\ref{eq:sixFun}) then the alternative
$b$--$\ell$ pairing is considered and the procedure is repeated.
If more than one solution is found, the one with
the minimum product of $\pt^{\nu}$ and $\pt^{\bar{\nu}}$ is selected.
About 70\% of signal $t\bar{t}$ simulated events
and 50\% of background events are reconstructed.

The number of expected and observed events in each channel
after selection and reconstruction is listed in Table~\ref{tab:eventYield}.

\newcommand\TT{\rule{0pt}{+2.3ex}}
\newcommand\BB{\rule[-1.5ex]{0pt}{0pt}}

\begin{table}[htbp]
 \caption{
The number of expected and observed events after reconstruction for the $\ee$, $\mumu$ and $\emu$ channels.
The combined statistical uncertainty and uncertainty due to the
cross-section are included for samples derived from MC simulation.
 }
 \begin{center}
   \begin{tabular}{l | r @{$\ \pm\ $}l| r @{$\ \pm\ $}l| r @{$\ \pm\ $}l}
   \hline
  Source  &   \multicolumn{2}{c|}{~~~$\ee$ channel~~~} &  \multicolumn{2}{c|}{~~~$\mumu$ channel~~~} & \multicolumn{2}{c}{~~~$\emu$ channel~~~}  \\
   \hline
   \hline
 $\ttbar$ signal               &~~~~~352 & 21      &~~~~1016 & 60  &~~~~2950 & 170  \\
  Single top ($Wt$-channel)     &9.0 & 0.9   &25.4 & 2.0 &90 & 7  \\
   Fake leptons                  & 4.8  &  2.2   & 8.0 & 2.8  &33 & 6   \\
\TT \BB $Z$ ($\rightarrow \ee$/$ \mumu$)+jets  &   0.6 & 0.4  &1.1 & 0.6  &  \multicolumn{2}{c}{~~~\textendash} \\
\TT \BB $Z$ ($\rightarrow \tautau$)+jets      &   0.32 & 0.29   &1.5 & 0.6  & 9.4 & 1.4  \\
\TT \BB Diboson                       &   0.25 & 0.08   &1.4 & 0.5  &5.4 & 1.3  \\
   \hline
 Total expected                &366 & 21      &1054 & 60  &3080 & 180  \\
   \hline
 Observed               &  \multicolumn{2}{c|}{383}            & \multicolumn{2}{c|}{1082}         &  \multicolumn{2}{c}{3132}  \\
   \hline
   \end{tabular}
 \end{center}

 \label{tab:eventYield}
\end{table}

The distribution of reconstructed $\coscos$ for the sum of the three dilepton channels,  
with the signal \ttbar\ simulated sample from {\sc MC@NLO} and backgrounds overlaid,
is shown in Figure~\ref{fig:ctpSC0}.
The backgrounds
are highly suppressed by the $b$-tagging requirement.
The expectation is in good agreement with data.

\begin{figure}[htpb]
  \centering
    \includegraphics[width=0.55\paperwidth]{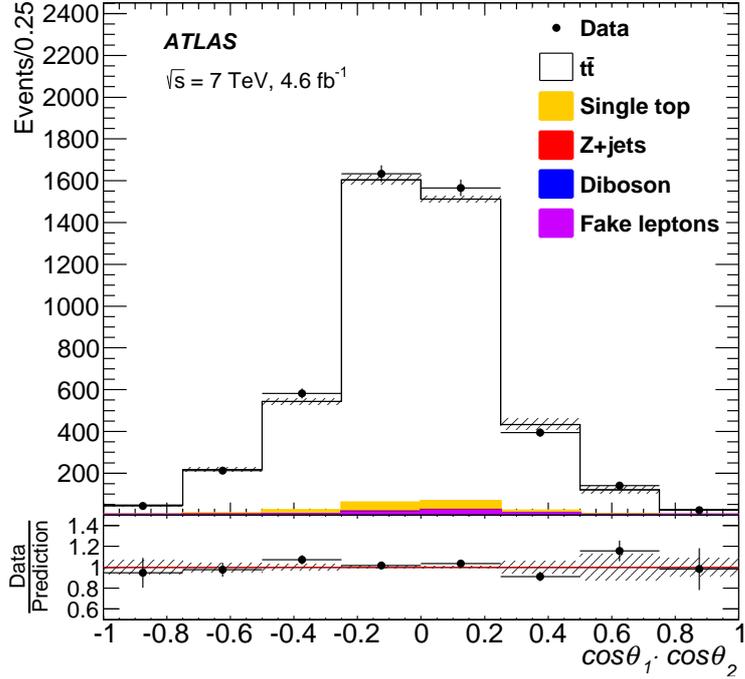}
  \caption{
The distribution of the reconstructed $\coscos$ for selected events.
The distribution for the signal \ttbar\ simulated sample with SM spin correlation
is overlaid and the estimated backgrounds
are highly suppressed due to the $b$-tagging requirement.
The hatching represents the total(stat.+syst.) uncertainty of the predictions.
  }
  \label{fig:ctpSC0}

\end{figure}

\section{Statistical method and validation}
\label{sec:unfolding}

\subsection{Unfolding method}

The distribution of $\coscos$ is distorted due to the resolution and acceptance of the detector.
An unfolding method is used to build an estimator for the $\coscos$ distribution at parton level
from the reconstructed distribution by correcting for such effects.

Prior to unfolding, the backgrounds listed in Table~\ref{tab:eventYield} are subtracted from data. 
The $\ttbar$ events where
one or both of the $W$ bosons decay to a $\tau$ that subsequently decays to an $e$ or $\mu$
are taken from simulation and are also subtracted from data.

The number of bins used in the unfolding is chosen based on
studies of the resolution of the $\coscos$ reconstruction and taking
into account the number of selected events in data,
while minimizing the bin-to-bin correlations.
Eight equally sized bins in $\coscos$ are used.

A true physical observable in bin $C_i$ of distribution $n(C_i)$ is related to the
reconstructed quantity in bin $E_j$ of a distribution $n(E_j)$ by the response matrix $ P(E_j \, | \, C_i)$,
which represents the event migration probability from bin $C_i$ to bin $E_j$:

\begin{equation}
\begin{split}
 n(E_j) &= \sum_{i=1}^{n_C} \, P(E_j \, | \, C_i) \, \epsilon_i \, n(C_i),~~~~~~j = 1, ..., n_E,
\end{split}
\label{eq:ecMatrix}
\end{equation}

\noindent
where $\epsilon_i$ represents the measurement efficiency of events in bin $C_i$
and $n_C$ and $n_E$ represent the total number of bins in the true and reconstructed distributions respectively.

In this analysis an iterative Bayesian unfolding method is used~\cite{bib:ufdHEP,bib:ufdBayes}, in which
Bayes' theorem is adopted to produce the following conditional probability:

\begin{equation}
P(C_i \, | \, E_j) = \frac{P(E_j \, | \, C_i) \, P_0(C_i)} {\sum_{l=1}^{n_C}P(E_j \, | \, C_l) \, P_0(C_l)},
\label{eq:invertM}
\end{equation}
\begin{equation}
\epsilon_i \equiv \sum_{j=1}^{n_E} P(E_j\, | \,C_i),
\label{eq:invertM2}
\end{equation}

\noindent
where $P(C_i \, | \, E_j)$ represents the
probability of having a true event in bin $C_i$, given a reconstructed event in bin $E_j$.
And $P_0(C)$ is the normalized \textit{prior}
distribution of $n(C_i)$. Using $P(C_i \, | \, E_j)$,
one can calculate $n(C)\epsilon_i$ using the following equation:

\begin{equation}
\hat{n}(C_i)\epsilon_i = \sum_{j=1}^{n_E} n(E_j) P(C_i \, | \, E_j).
\label{eq:nUfd}
\end{equation}

\noindent
The probability $P(C_i \, | \, E_j)$ depends on the \textit{prior} distribution of $P_0(C)$,
and $\hat{n}(C_i)\epsilon_i$ is a biased estimator if $P_0(C)$ differs from data.
To reduce the bias, an iterative procedure is introduced,
replacing $P_0(C)$ with the normalized $\hat{n}(C_i)\epsilon_i$
in Eq.(~\ref{eq:invertM}) to recalculate $P(C_i \, | \, E_j)$ and
then $\hat{n}(C_i)\epsilon_i$ using Eq.(~\ref{eq:nUfd}).
Finally, $\hat{n}(C_i)$ is obtained by scaling $\hat{n}(C_i)\epsilon_i$ using $1/{\epsilon_i}$ from MC simulation.

Increasing the number of iterations reduces the bias of the estimator.
However, fluctuations and correlations between bins of the estimator are increased.
The iterative procedure is repeated until the unfolded distribution
in the current iteration is consistent with the unfolded distribution in the previous iteration
within the statistical uncertainty.  In this case, further iterations would not increase the sensitivity.
Therefore the termination criterion is defined as:

\begin{equation}
\frac{\chi^2}{N_{\text{bins}}} \leq 1,
\label{eq:chi2OverBins}
\end{equation}

where

\begin{equation}
\begin{split}
\chi^2 = &\sum_{j=1}^{N_\text{bins}}~\sum_{i=1}^{N_\text{bins}}\left(n_{i}' - n_{i}\right)
         \left(\sigma_{i,j}\right)^{-1} \left(n_{j}' - n_{j}\right)^{\rm T},
\end{split}
\label{eq:chi2Def}
\end{equation}

in which $\left(n_{i}' - n_{i}\right)$
refers to the $i$th bin difference between the unfolded distribution
$n_{i}'$ in the current iteration and the unfolded distribution $n_{i}$ in the previous iteration,
$\sigma_{i,j}$ is the covariance matrix of the unfolded distribution $n_{i}'$,
and $N_\text{bins}$ is the number of bins in the unfolded distribution.
The larger the
difference between the \textit{prior} distribution and the real distribution, the more iterations are required.
Studies using MC simulation have shown that three iterations suffice
to reduce the bias below the level of the statistical uncertainty.

\subsection{Method validation}

The unfolding method is validated and its uncertainty obtained using a MC sample containing the SM spin correlation.
The simulated \coscos\ distribution after detector simulation, selection, and reconstruction is 
compared to that in collision data, and the ratio of the two is fitted by a smooth function that
is used to weight the parton-level distribution.  This is propagated through to the reconstructed distribution and 
results in a pseudomeasurement and a corresponding parton-level distribution.
The unfolding method, using the nominal response matrix, is applied to the pseudomeasurement.
The systematic uncertainty on the unfolding method is taken as the difference between the 
unfolded pseudomeasurement and the known parton-level distribution and is shown in Table~\ref{tab:syst_dilep_s}.

\section{Systematic and statistical uncertainties}
\label{sec:systDescription}

\subsection{Systematic uncertainties}

Systematic uncertainties are evaluated by applying the unfolding
procedure (using the nominal unfolding matrix) to pseudoexperiments
created using MC samples modified
to reflect the various systematic uncertainties.
The systematic uncertainty of the unfolded distribution is then obtained by
comparing the varied unfolded distributions to the nominal unfolded distribution.
The following systematic uncertainty sources are considered in this analysis.

\textbf{MC generator modeling:}
The uncertainty due to generator modeling is assessed using three different groups of samples.
{\sc Powheg+Pythia}~\cite{Nason:2004rx, Frixione:2007vw, Alioli:2010xd, Sjostrand:2006za} is compared to {\sc MC@NLO+Herwig}, where both the generator and parton showering are varied.
{\sc Powheg+Pythia} is compared to {\sc Alpgen+Herwig} and finally {\sc Powheg+Pythia} is compared to {\sc Powheg+Herwig}, where only
the parton showering is different.
The largest variation of the unfolded distributions found in these three comparisons is taken as the uncertainty.

\textbf{ISR/FSR:}
The uncertainty due to initial-state radiation / final-state radiation (ISR/FSR) is evaluated using
{\sc Alpgen+Pythia} samples within which the parameters controlling ISR and FSR are varied within
a range consistent with data.  The average of the absolute values of the upwards and downwards variations of the unfolded distributions
is taken as the systematic uncertainty.

\textbf{PDF:}
The impact of the choice of PDF in
simulation was studied by reweighting the MC samples to three PDF sets
(CT10, MSTW2008~\cite{Martin:2009iq} NLO, and NNPDF20~\cite{Ball:2010de})
and taking half of the maximum
difference of the unfolded distributions using any two PDF sets.

\textbf{Underlying event and color reconnection:}
To estimate the effect of the underlying event (UE), two samples simulated by {\sc Powheg+Pythia}
with the Perugia 11 and Perugia 11 mpiHI sets of tuned parameters are used~\cite{PhysRevD.82.074018}.
The variation of the unfolded distributions between these two tunes is taken as the systematic uncertainty.
The impact due to the modeling of color reconnection is studied by comparing two samples simulated with {\sc Powheg+Pythia}.
One has the nominal color reconnection model
and the other has no color reconnection.
The difference of the unfolded distributions between these two samples is taken as the systematic uncertainty.

\textbf{Jet energy scale and jet reconstruction:}
The relative jet energy scale (JES) uncertainty varies from 1\% to 3\% depending on jet $\pt$ and $\eta$~\cite{PERF-2012-01}.
The jet reconstruction efficiency for data and the MC simulation is found to be in agreement with an accuracy of better than $\pm2\%$~\cite{PERF-2011-03}.
To account for the residual uncertainties, 2\% of jets with $\pt < 30$~\GeV\ are randomly removed from MC simulated events.
The uncertainty related to the JVF is less than 1\% and depends on jet $\pt$.
For all jet-related systematic uncertainties the differences are propagated to the unfolded distribution and the variation taken as the uncertainty.

\textbf{$b$-tagging efficiency:}
Differences in the $b$-tagging efficiency as well as $c$-jet and light-jet mistag rates in data and simulation are parameterized
using correction factors, which are functions of \pt\ and $\eta$~\cite{ATLAS-CONF-2012-040}.
The uncertainty on these correction factors is propagated to the unfolded distribution.

\textbf{Modeling of \met:}
Uncertainties on the energy scale of jets and leptons are also propagated to the
uncertainty on \met. Other contributions to this uncertainty originate from the energy scale and
resolution of the soft calorimeter energy deposits that are not included in the reconstructed jets
and leptons, and is propagated to the uncertainty of the unfolded distribution.

\textbf{Lepton reconstruction:}
The modeling of the lepton momentum scale and resolution is studied using the
reconstructed dilepton invariant mass distribution of $Z \rightarrow \ell^+\ell^-$ candidates
and the simulation is adjusted accordingly.
Any mismodeling of the electron and muon trigger, reconstruction, and selection
efficiencies in the simulation is corrected using
measurements of the efficiency in data.
The systematic uncertainties on the correction factors applied are propagated to the unfolded distribution.

\textbf{Luminosity uncertainty:}
The uncertainty on the measured integrated luminosity is 1.8\%~\cite{DAPR-2011-01}.
The effect of the luminosity uncertainty is evaluated by scaling the number of signal and background
events by the luminosity uncertainty, for processes estimated exclusively from simulation.
The change in the result due to the luminosity uncertainty is taken as systmatic uncertainty.

\textbf{Background uncertainties:}
The uncertainties due to the normalization of the nonprompt and fake lepton estimate, $Wt$ channel single top,
$Z$+jets, and diboson events are propagated to the uncertainty of the unfolded distribution.

\textbf{Bayesian unfolding method:}
The residual bias in the unfolding method is taken as a systematic uncertainty as described in Section~\ref{sec:unfolding}.

The
evaluated systematic uncertainties are listed in Table~\ref{tab:syst_dilep_s}, for each bin of the $\coscos$ distribution.
The result of varying the top quark mass by $\pm1$~\GeV\
is shown in the last row of the table.
The main sources of uncertainty are the unfolding method,
followed by the uncertainties associated with jets.
Some theoretical uncertainties (MC generator, top quark mass, UE/color connection)
are estimated with uncorrelated MC samples, and hence include a statistical uncertainty
due to the MC sample size.

\begin{table}[htbp]
\caption{
Relative uncertainties (in \%) for each bin of the
normalized unfolded $\coscos$ distribution.
Where the magnitudes of the upwards and downwards systematic uncertainties differ,
the larger of the two is taken. 
The `Total'  shows the sum in quadrature of the individual components.
}
\begin{center}
\resizebox{\textwidth}{!} { 
  \begin{tabular}{| c | c | c | c | c | c | c | c | c |}
  \hline
  Bin range &$-1$ : $-0.75$ &$-0.75$ : $-0.5$ &$-0.5$ : $-0.25$ &$-0.25$ : 0 &0 : 0.25 &0.25 : 0.5 &0.5 : 0.75 &0.75 : 1 \\
  \hline
Generator modeling		&6.9 	&3.2 	&1.6 	&0.5 	&0.8 	&2.2 	&1.0 	&0.0\\
ISR/FSR	  	&2.0 	&0.9 	&0.6 	&0.3 	&0.3 	&1.1 	&1.0 	&0.8\\
PDF			&0.5 	&0.3 	&0.1 	&0.0 	&0.0 	&0.2 	&0.2 	&0.0 \\
UE/color reconnection			&1.5 	&1.1 	&1.0 	&0.7 	&0.1 	&0.5 	&0.6 	&3.1\\
\hline	
JES/jet reconstruction			&4.5 	&3.0 	&1.1 	&0.6 	&0.9 	&1.1 	&1.8 	&3.1 \\
$b$-tagging SF			&0.0 	&0.3 	&0.0 	&0.1 	&0.0 	&0.1 	&0.2 	&0.0\\
$\met$			&0.5 	&0.6 	&0.4 	&0.1 	&0.1 	&0.3 	&0.2 	&0.0\\
Lepton reconstruction			&1.5 	&0.6 	&0.1 	&0.3 	&0.1 	&0.5 	&0.6 	&0.8 \\
Luminosity uncertainty		&0.5 	&0.1 	&0.0 	&0.1 	&0.0 	&0.1 	&0.2 	&0.0 \\
Background uncertainty		&1.5 	&0.6 	&0.4 	&0.1 	&0.1 	&0.4 	&0.6 	&0.8 \\
Bayesian unfolding method		&10.9 	&0.6 	&2.3 	&1.4 	&1.0 	&2.6 	&0.6 	&7.8\\
\hline	
Total			&13.9 	&4.9 	&3.3 	&1.8 	&1.7 	&3.9 	&2.7 	&9.3\\
\hline	
Top quark mass ($\pm1$~\GeV)	& 0.1 	& 0.2 	& 0.1 	& 0.2 	& 0.1 	& 0.3 	& 0.0 	& 0.6\\
   \hline
   \end{tabular}
}
\end{center}

\label{tab:syst_dilep_s}
\end{table}

\subsection{Statistical uncertainties}
\label{sec:staDescription}

The uncorrelated bin-to-bin statistical uncertainties of the $\coscos$ distribution at reconstruction
level are propagated to the unfolded distribution, in which bin-to-bin correlations arise.
The $8\times8$ correlation matrix is shown in Table~\ref{tab:correlationFactor}.

\begin{table}[htbp]
\caption{The correlation factors for the statistical uncertainties between any two bins of the unfolded distribution.}
\begin{center}
  \begin{tabular}{ | c | c c c c c c c c |}   \hline
  Bin number & 1 & 2 & 3 & 4 & 5 & 6 & 7 & 8 \\
  \hline
  1  & 1       &           &           &         &          &        &       &       \\
  2  & 0.74    &       1       &       &       &       &       &          &         \\
  3  & 0.29    &       0.68    &       1       &           &         &          &          &          \\
  4  & $-$0.026  &       0.071   &       0.44    &       1       &           &          &          &         \\
  5  & $-$0.13   &       $-$0.18   &       $-$0.062  &       0.58    &       1       &           &           &           \\
  6  & $-$0.093  &       $-$0.14   &       $-$0.16   &       0.049   &       0.59    &       1       &           &           \\
  7  & $-$0.068  &       $-$0.13   &       $-$0.18   &       $-$0.08   &       0.28    &       0.75    &       1       &           \\
  8  & $-$0.035  &       $-$0.081  &       $-$0.13   &       $-$0.083  &       0.14    &       0.49    &       0.8     &       1       \\
  \hline
  \end{tabular}
\end{center}
\label{tab:correlationFactor}
\end{table}

The statistical and systematic uncertainties for each bin of the  
unfolded $\coscos$ distribution is summarized in Table~\ref{tab:listCmbDiStatSyst}.

\section{Results}
\label{sec:DifResult}

The unfolded distribution of $\coscos$ is shown in Figure~\ref{fig:difChnStatSyst} and presented in Table~\ref{tab:listCmbDiStatSyst}. 
The distribution is compared to the prediction from {\sc MC@NLO} giving a $\chi^2/N_{\text{bin}} = 4.1/8$.
Individual analyses for the $\ee$, $\mumu$, and $\emu$ channels are performed 
and the measurements are found to be consistent with the combined result.

\begin{figure}[htpb]
  \centering
  \includegraphics[width=0.6\paperwidth,height=0.5\paperwidth]{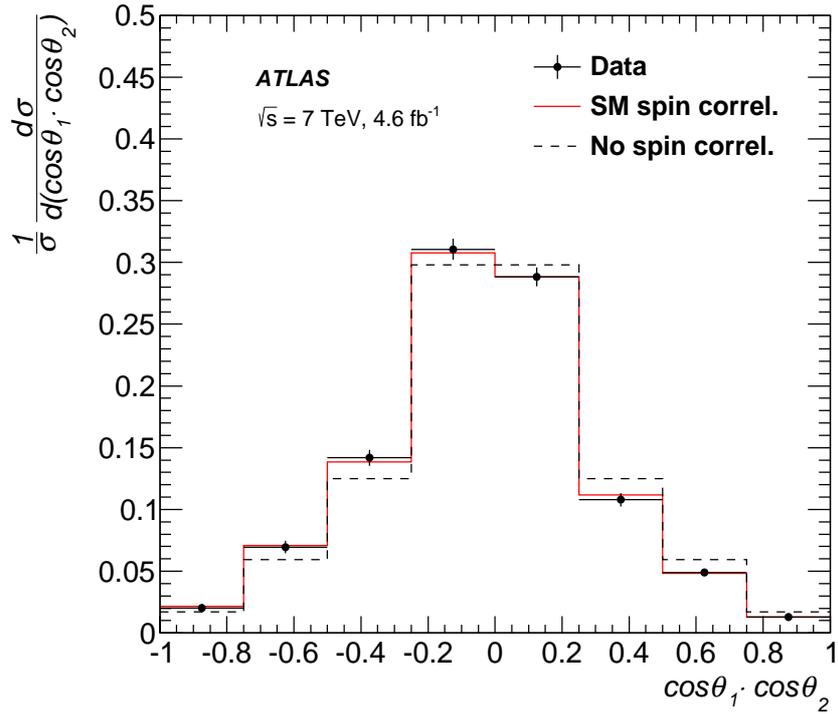}
  \caption{
The unfolded data distribution of $\coscos$, including the statistical and systematic uncertainties summed in quadrature.
The predictions from SM and the {\sc MC@NLO} sample without spin correlation are overlaid for comparison.
A symmetric distribution around zero would indicate no spin correlation.
}
  \label{fig:difChnStatSyst}
\end{figure}

\begin{table}[htbp]
\caption{
The numerical summary of the unfolded $\coscos$ distribution, with statistical and systematic uncertainties.
The SM prediction is shown in the last column for comparison.
}
\begin{center}
  \begin{tabular}{ r@{~~:~~}l | c | c }   \hline
  \multicolumn{2}{c|}{Bin range}   & Unfolded data & {\sc MC@NLO} prediction \\  
  \multicolumn{2}{c|}{}   &  $1 / \sigma~~\textrm{d}\sigma / \textrm{d}(\coscos)$  $\pm$ stat. $\pm$ syst.  & $1 / \sigma~~\textrm{d}\sigma / \textrm{d}(\coscos)$ $\pm$ stat. \\  
  \hline  \hline 
  $-$1.00  & $-$0.75  & $0.0202\pm0.0020\pm0.0028$ & $0.0215\pm0.0005$ \\
  $-$0.75  & $-$0.50  & $0.0696\pm0.0037\pm0.0034$ & $0.0707\pm0.0008$ \\
  $-$0.50  & $-$0.25  & $0.1418\pm0.0045\pm0.0047$ & $0.1384\pm0.0010$ \\
  $-$0.25  & 0      & $0.3106\pm0.0062\pm0.0057$ & $0.3079\pm0.0014$ \\
  0      & 0.25   & $0.2882\pm0.0059\pm0.0048$ & $0.2884\pm0.0013$ \\
  0.25   & 0.50   & $0.1078\pm0.0033\pm0.0042$ & $0.1118\pm0.0009$ \\
  0.50   & 0.75   & $0.0489\pm0.0024\pm0.0013$ & $0.0484\pm0.0006$ \\
  0.75   & 1.00   & $0.0129\pm0.0009\pm0.0012$ & $0.0129\pm0.0003$\\  \hline
  \end{tabular}
\end{center}
\label{tab:listCmbDiStatSyst}
\end{table}

Previous publications quote the results in terms of 
$A_{\textrm{helicity}} = (N_{\textrm{like}} - N_{\textrm{unlike}})/(N_{\textrm{like}} + N_{\textrm{unlike}})$
where $N_{\textrm{like}}$ ($N_{\textrm{like}}$) is the number of events where the 
top quark and top antiquark have parallel (anti-parallel) spins with respect to the helicity basis.
To compare with these quantitatively, the parameter $C_{\textrm{helicity}}$ in Eq.(~\ref{eq:spinAngDisTTbar})
is extracted from the unfolded distribution using $C_{\textrm{helicity}}  = -9 \left< \coscos \right>$~\cite{Bernreuther:2010ny,Bernreuther:2015yna}.
This is converted to $A_{\textrm{helicity}}$ using $C_{\textrm{helicity}} =-A_{\textrm{helicity}} \alpha_1 \alpha_2$, where $\alpha_1$ and $\alpha_2$ 
are the spin analyzing powers for the two charged leptons as in Ref.~\cite{TOPQ-2013-01}.
In dilepton final states the spin-analyzing power is effectively 100\%, therefore $C=A$.
This results in $A_{\textrm{helicity}} = 0.315 \pm 0.061 \textrm{(stat.)} \pm 0.049 \textrm{(syst.)}$,
which agrees well with the NLO QCD prediction of $A_{\textrm{helicity}} =0.31$~\cite{Bernreuther:2001rq},
the previous measurements using 
template fits to event properties without correcting for detector acceptance and efficiencies by ATLAS~\cite{TOPQ-2011-11,TOPQ-2013-01,TOPQ-2014-07},
and the unfolded parton level results reported by CMS~\cite{CMS-TOP-13-003}.

\section{Conclusion}
\label{sec:conclusion}

A differential cross-section measurement of the \coscos\ distribution is presented using 4.6~fb$^{-1}$ of
proton--proton collision data
collected at $\rts = 7$~\TeV\ by the ATLAS detector at the LHC during 2011\@.
Events are selected in the dilepton topology with two jets.
The background rejection is improved by the use of $b$-tagging.
The distribution of $\coscos$ is reconstructed using the kinematic information about
the selected objects and unfolded to parton level using
an iterative Bayesian unfolding algorithm.
The unfolded distribution
is in good agreement with the prediction from {\sc MC@NLO}\@.
The main sources of uncertainty are due to the unfolding method, 
theoretical modeling of the signal, and uncertainties related to the reconstruction of jets.

\section*{Acknowledgments}
We thank CERN for the very successful operation of the LHC, as well as the
support staff from our institutions without whom ATLAS could not be
operated efficiently.

We acknowledge the support of ANPCyT, Argentina; YerPhI, Armenia; ARC,
Australia; BMWFW and FWF, Austria; ANAS, Azerbaijan; SSTC, Belarus; CNPq and FAPESP,
Brazil; NSERC, NRC and CFI, Canada; CERN; CONICYT, Chile; CAS, MOST and NSFC,
China; COLCIENCIAS, Colombia; MSMT CR, MPO CR and VSC CR, Czech Republic;
DNRF, DNSRC and Lundbeck Foundation, Denmark; EPLANET, ERC and NSRF, European Union;
IN2P3-CNRS, CEA-DSM/IRFU, France; GNSF, Georgia; BMBF, DFG, HGF, MPG and AvH
Foundation, Germany; GSRT and NSRF, Greece; RGC, Hong Kong SAR, China; ISF, MINERVA, GIF, I-CORE and Benoziyo Center, Israel; INFN, Italy; MEXT and JSPS, Japan; CNRST, Morocco; FOM and NWO, Netherlands; BRF and RCN, Norway; MNiSW and NCN, Poland; GRICES and FCT, Portugal; MNE/IFA, Romania; MES of Russia and NRC KI, Russian Federation; JINR; MSTD,
Serbia; MSSR, Slovakia; ARRS and MIZ\v{S}, Slovenia; DST/NRF, South Africa;
MINECO, Spain; SRC and Wallenberg Foundation, Sweden; SER, SNSF and Cantons of
Bern and Geneva, Switzerland; NSC, Taiwan; TAEK, Turkey; STFC, the Royal
Society and Leverhulme Trust, United Kingdom; DOE and NSF, United States of
America.

The crucial computing support from all WLCG partners is acknowledged
gratefully, in particular from CERN and the ATLAS Tier-1 facilities at
TRIUMF (Canada), NDGF (Denmark, Norway, Sweden), CC-IN2P3 (France),
KIT/GridKA (Germany), INFN-CNAF (Italy), NL-T1 (Netherlands), PIC (Spain),
ASGC (Taiwan), RAL (UK) and BNL (USA) and in the Tier-2 facilities
worldwide.

\printbibliography

\newpage
\begin{flushleft}
{\Large The ATLAS Collaboration}

\bigskip

G.~Aad$^\textrm{\scriptsize 85}$,
B.~Abbott$^\textrm{\scriptsize 113}$,
J.~Abdallah$^\textrm{\scriptsize 151}$,
O.~Abdinov$^\textrm{\scriptsize 11}$,
R.~Aben$^\textrm{\scriptsize 107}$,
M.~Abolins$^\textrm{\scriptsize 90}$,
O.S.~AbouZeid$^\textrm{\scriptsize 158}$,
H.~Abramowicz$^\textrm{\scriptsize 153}$,
H.~Abreu$^\textrm{\scriptsize 152}$,
R.~Abreu$^\textrm{\scriptsize 116}$
% ...
\bigskip
\\
% ...
$^{11}$ Institute of Physics, Azerbaijan Academy of Sciences, Baku, Azerbaijan\\
% ...
$^{85}$ CPPM, Aix-Marseille Universit{\'e} and CNRS/IN2P3, Marseille, France\\
% ...
$^{90}$ Department of Physics and Astronomy, Michigan State University, East Lansing MI, United States of America\\
% ...
$^{107}$ Nikhef National Institute for Subatomic Physics and University of Amsterdam, Amsterdam, Netherlands\\
% ...
$^{113}$ Homer L. Dodge Department of Physics and Astronomy, University of Oklahoma, Norman OK, United States of America\\
% ...
$^{116}$ Center for High Energy Physics, University of Oregon, Eugene OR, United States of America\\
% ...
$^{151}$ Institute of Physics, Academia Sinica, Taipei, Taiwan\\
$^{152}$ Department of Physics, Technion: Israel Institute of Technology, Haifa, Israel\\
$^{153}$ Raymond and Beverly Sackler School of Physics and Astronomy, Tel Aviv University, Tel Aviv, Israel\\
% ...
$^{158}$ Department of Physics, University of Toronto, Toronto ON, Canada\\
% ...
\end{flushleft}


\end{document}